# Performance Dependency of LSTM and NAR Beamformers With Respect to Sensor Array Properties in V2I Scenario

**Prateek Bhadauria[1], Ravi Kumar*[1], Sanjay Shamra[1]**

bhadauria.prateek@gmail.com, roybhu_royravs@yahoo.co.in, sanjay.sharma@thapar.edu

[1]Thapar Institute of Engineering and Technology, Patiala, Punjab, India.

Department of Electronics and Communication

## Summary

Prediction and nullifying the interference is a challenging problem in vehicle to infrastructure (V2I) scenarios. The implementation of practical V2Inetwork is limited because of inevitability of interference due to random nature of the wireless channel. The interference introduces angle ambiguity between the road side unit's (RSU's) mounted base station and user equipment (UE). This paper proposes an adaptive beamforming (ABF) technique for mitigation of interference in V2I networks, especially in a multiuser environment. In this work, LSTM based deep learning and Non-Linear Auto Regressive (NAR) based regressor have been employed to predict the angles between the RSU's and UE. Advance prediction of transmit and receive signals enables reliable V2Icommunication. Instead of predicting the beamforming matrix directly, we predict the main features using LSTM for learning dependencies in the input time series where complex variables were taken as input states and final beamformed signal was the output. Simulation results have confirmed that the proposed LSTM model achieves comparable performance in terms of system throughput when compared with the non-linear auto regressive(NAR) method implemented as an artificial neural network (ANN).

## 1 | INTRODUCTION

The fifth-generation (5G) networks have opened up new avenues on a plethora of vehicle to infrastructure(V2I) scenarios. With the utilization of mmWave band, it is possible to make compact design of antenna providing more directional beams towards the UE. Although the radio waves interacting with the propagation environment create interference generated due to the blockage caused by the buildings; mm wave frequencies influence network coverage of a cell significantly.[1] Furthermore, due to smaller wavelengths, blockages by various objects like vehicles, vegetation, urban furniture and even human bodies become significant. Although different schemes and models are available to mitigate the interference, thanks to the adaptive beamforming (ABF) techniques (also referred to as the smart antenna technique) closely spaced antenna arrays (e.g., half-wavelength spacing) with sufficient antenna correlation are able to improve the performance of a system by predicting the direction of arrival (DOA). In this context, much of the research work has been carried out for acoustic signals and different conventional algorithms applied in the field of medical imaging, naval communication, speech processing etc.[2-4]

Of late, there have been varied efforts for implementing an adaptive beamformer in various network scenarios. In Toch et al,[5] a deep network was implemented initially to recover the representation of users in the social network and subsequently a basic recurrent neural network (RNN) and long short-term memory (LSTM) model were used to explore mobility configurations of a single user at different levels. In Yang et al,[6] an online method of learning was investigated to train a convolutional neural network (CNN) based on hierarchical learning, in which streams of data were processed in parallel. However, previous works have by and large focused on prediction of individual trajectories rather than forecasting network-level user distribution, which in fact is of a greater value. It was felt that, we can directly predict the user density at network level in spatial domain through a deep learning (DL)approach. Since fluctuation of signals within the channel conditions is influenced by short and long term behaviors, the LSTM network, a kind of deep RNN, comes across as a suitable candidate to predict the number of UEs in every limited area due to its inherent quality by learning long-term dependencies. By using the LSTM neural network, the applied technique overcomes the vanishing gradient problem that creates ambiguities in conventional RNNs and captures long-short term temporal and spatial dependencies without suffering from optimization hurdles.[7] The array response pattern of received signal depends upon proper adjustment of beamforming weights which can eventually mitigate the interference .adaptive beamforming algorithm has also been proved effective for V2I networks in terms of mitigating the interference, johnson et al[8] studied the probabilistic approach to quantify the channel uncertainty using adaptive beam steering algorithm, Vouras et al[9] investigated the interference cancellation techniques by applying adaptive algorithms for high frequency radar. Montesinos et al[10] adopted a method for beamforming with the use satellite multiple beam antennas (MBA) which signifies the important role of ABF. To implement adaptive beamforming it is necessary to steer the beam and update the weights with the dynamic environment and to develop the techniques helpful in rejection of interference. Therefore several co-channel interference rejection techniques have been formulated for cellular communication as reported in,[11] Bartlett beamformer, a data independent method, has used as a conventional beamforming technique.[12] It maximizes the power at the output of the desired signal, which gives a spatially matched filtering to the signal of interest (SOI). Since the received signals are independent of data and its capability for mitigation of interference is relatively limited. In today's scenario data dependent methods of beamforming are used more prolifically e.g., the Capon beamformer. Capon beamformer,[12] keeps the array gain of the SOI unity while minimizing the second-order statistics (SOS) of the output. Minimum-variance distortionless response (MVDR) or Capon's method gives a reliable output to predict and estimate the DOA, As an adaptive technique, it is a promising approach for extracting the features to map the underlying function between input and output sets.[13] In addition to it, by accomplishing hierarchical extraction of features, the Capon's method is capable of obtaining both spatial and temporal frameworks in sequential data, while minimizing the hand crafted data pre-processing tasks.[14] The Capon method estimates power spectral density of a time series by employing an FIR filter that suppresses all but a particular frequency of an input signal.

The MVDR beamformer has much better resolution and interference suppression capability than conventional data-independent beamformers. As the usage of data dependent techniques becomes more prominent the Meng et al[3] formulated a deep learning (DL) technique which is applicable for beam management, beam selection and collected data includes (e.g., vehicle's position). The prediction of traffic in heterogeneous network model uses real-time data collected by different types of road structure mounted with cameras or sensors. The predominant model for prediction of traffic is most common which uses time series models, mostly depends on the past values of traffic to forecast the next one.

In the backdrop of adaptive learning techniques and their inherent capabilities, RNNs provide a unique alternative for deployment of machine learning for allowing perceptual deviations and predicting dense traffic by representing the inputs as time series data.[14] Nevertheless, these RNNs which are traditional in nature, such as simple RNN and gated recurrent unit (GRU), are not capable to memorize the long-term dependencies in different scenarios ofV2I used cases of 5G applications due to unusual behavior of pedestrians and vehicles on the road.[15,16] Subsequently, a specific type of architecture of RNN, the LSTM have implemented to focus on these constraints in context of time series prediction.[17]

During the last couple of years, the LSTM networks have been implemented successfully and find applications in robotics, speech processing, natural language processing, traffic prediction via data collected from sensors mounted on road side units (RSU's). [18,19] Apart from evaluating RNNs in the form of LSTM, we also examine the use of the nonlinear autoregressive (NAR) neural network method as a prediction technique for interference mitigation. Forecasting the interference time series signal, that are very noisy and nonstationary, is a challenging problem for V2I networks in 5G. In this context both NAR and LSTM based approaches are likely to give the output based on the past values of inputs. This possibility has been explored in our present work.

## 2 | CONTRIBUTION

The key contributions to this work can be summed up as follows:
- We apply a novel technique for prediction of beamformed signal in V2Iscenario using based MVDR beamformer and a neural network based a time series forecasting stage.
- It has been established through simulation experiments that the beamforming as a time series prediction problem can be successfully implemented using LSTM and NAR networks.
- In this paper time delay MVDR beamforming algorithm is used to steer the beam in the desired direction and applies a finite impulse response (FIR) filter at the output of each sensor element.
- The state of art of efficient ABF resolved the problem of mismatching the spatial signature for the non-stationary user that creates the interference inV2I based 5G scenario's

## 3 | SYSTEM MODEL

We have used a narrow band beamforming model for V2I scenario, and adopted a multiantenna setup where each BS comprises of uniform linear array (ULA) of $N$ antennas and user elements (UEs) signals assumed as a beamformer having single antenna, Thus the received signal at the UE from the output of beamformer is given by

$$R(t) = V^H X(t) \tag{1}$$

Where t is time step of the rays impinging at the UE sensor array $[X(t) = X_1(t), \ldots\ldots\ldots, X_N(t)]^T$ is the Nx1 snapshot vector of complex values, $V = [v_1, \ldots v_N]^T$ which is considered as a weight vector of complex variables (.)H and(.)T and denotes the Hermitian transpose and transpose respectively.

$$X(t) = C_d(t) + I(t)n(t) \tag{2}$$

Where $C_d(t), I(t)$, and $n(t)$ are the statistically independent variables of the desired user signal, interference and noise respectively. In this model slow fading case is applicable due to coherence time of the channel is large relative to the delay requirement of the V2I network. In this regime, the amplitude and phase change imposed by the channel can be considered roughly constant over the period of use so the vector $C_d(t)$ can be formulated as

$$C_d(t) = C(t)b_s \tag{3}$$

Where $C(t)$ is the waveform of complex variables of the desired signal and $b_s$ is its Nx1 spatial transmitted symbols which represents the wavefront of user, in such condition equation (2) can be rewritten as

$$X(t) = C(t)b_s + I(t) + n(t) \tag{4}$$

To determine the optimal beamforming vector at the user side, it is necessary to maximize the signal to interference plus noise ratio (SINR).[18]

$$\text{SINR} = \frac{V^H A_s V}{V^H A_{i+n} V} \tag{5}$$

Where $\quad A_s = E\{C_d(t)C_d^H(t)\}$

$$A_{i+n} = E\{I(t) + n(t)(I(t) + n(t)^H \tag{6}$$

Are called $N \times N$ signal and interference plus noise covariance metrices respectively and $E\{.\}$ represents expectation value used in statistics, normally $A_s$ is assumed as rank matrix with arbitrary values depends upon how fast the fading signal is, i.e., $1 \le \text{rank }\{A_s\} \le N$.

Consider a slow fading desired user signal then

$$A_s = \sigma_s^2 b_s b_s^H \tag{7}$$

Where $\sigma_s^2 = \{E|C(t)|^2\}$ in this case rank $\{A_s\} = 1$ and equation (5) becomes

$$\text{SINR} = \frac{\sigma_s^2 |V^H b_s|^2}{V^H A_{i+n} V} \tag{8}$$

By maximizing the SINR we can obtain the optimal weight vector in equation (5), The optimization problem is to maintain the array with property of distortionless response to formulate the beamformed signal while minimizing the output interference plus noise power.

$$\min_{w} V^H A_{i+n} V \text{ subject to } V^H A_s V = 1 \tag{9}$$

for the case of one rank signal, this condition can be rewrite in more sophisticated form [18-19].

$$\min V^H A_{i+n} V \text{ subject to } V^H b_s = 1 \tag{10}$$

This concept is generally called minimum variance distortionless response (MVDR) beamforming. The solution to equation (9) is given by the following generalized eigen value problem

$$A_{i+n} V = \lambda A_s V \tag{11}$$

Where $\lambda$ can be referred assign value that belongs to equation (11) and are non-negative real numbers that corresponds to positive semidefinite of $A_{i+n}$ and $A_s$.

Now the optimal weight can be formally referred as

$$V_{opt} = \mathcal{P}\{A_{i+n}^{-1} A_s\} \tag{12}$$

Where $\mathcal{P}(.)$ is the principal eigenvector operator of matrix equation (12) can be simplified for rank one signal source is presented as

$$V_{opt} = P\{A_{i+n}^{-1} b_s b_s^H\} = \beta A_{i+n}^{-1} b_s \tag{13}$$

Where the constant can be achieved from the MVDR constraint $v_{opt}^H b_s = 1$, in equation (10) and is equal to $\beta$.[20,21] In practical scenarios, the exact values of matrix $A_{i+n}$ is not achieved due to random nature of channel but it can be estimated by applying assumptions, so that the estimated values correspond to the optimization problems equation (9) and equation (10), its estimate should be used in optimization problems equation (9) and equation (10) than exact value.

As the reason to use adaptive beamforming is usually for canceling interference and it would be reasonable to base the criterion for optimization for the power at the output of the beamformer such that it is reduced by weight-set solution.

## 4 | PROPOSED METHODOLOGY

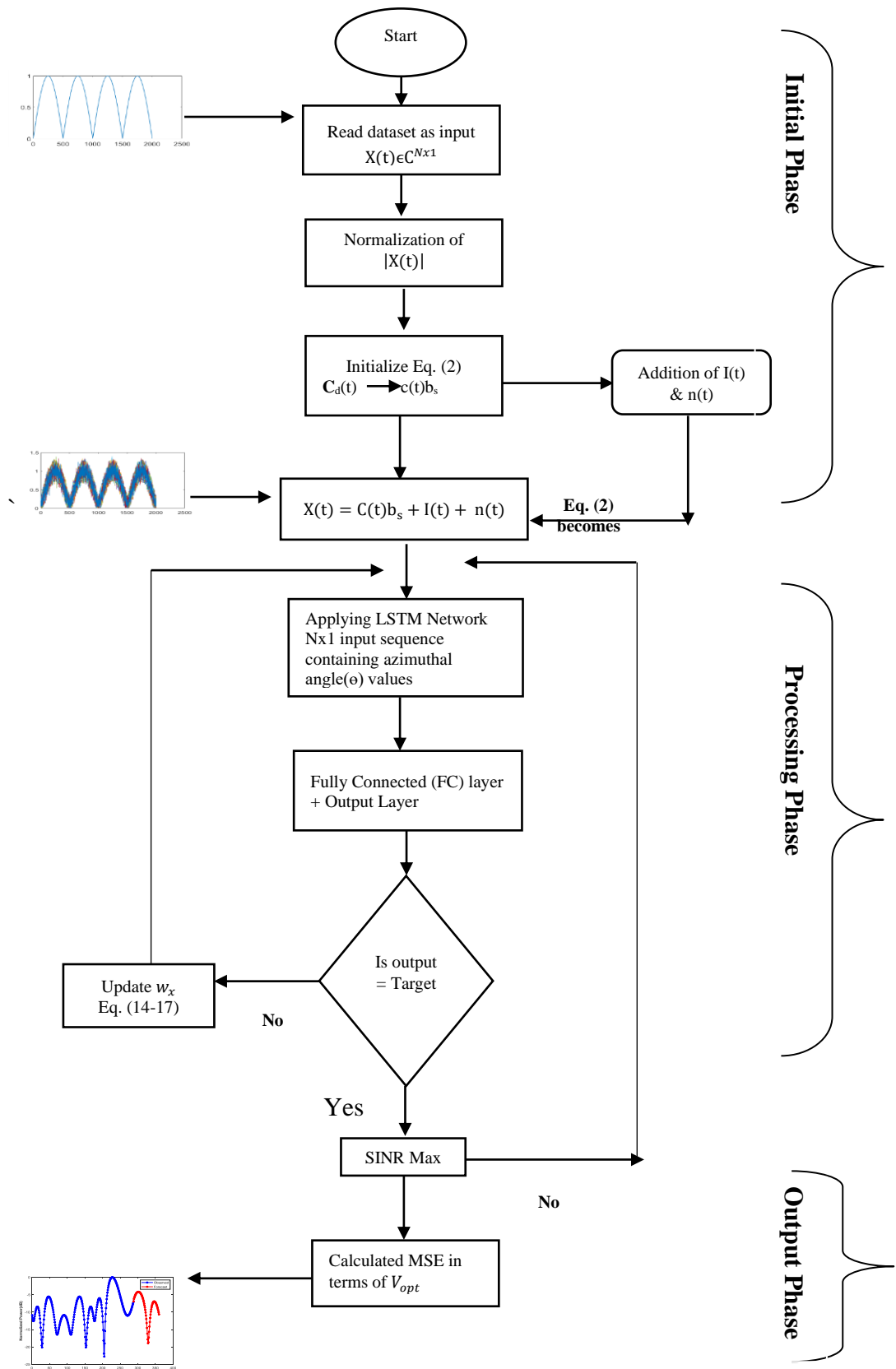

**FIGURE 1** The Proposed LSTM Model

Figure1 represents the various steps of the proposed methodology for the beamforming prediction. In the initial stage data is generated according to the given scenario discussed in section 5. The generated variables are complex in nature so it is necessary to be normalized for its further processing. Here the raw signal data is considered by adding some noise in it. Subsequently in the processing phase the signal is applied to the LSTM/NAR network (in the flow chart only LSTM has been indicated) and then fully connected layers are added for further processing where it learns and trains the weights when the input parameters contain azimuthal and incident angles and number of antenna elements. This stage is crucial and somewhat critical due to the parameters setting of the networks which includes the number of hidden layers, learning rate, dropout factor and amount of data is used for training and testing purpose, small variations in these parameters results drastic change in the output. Because all are dependent to each other in some respects within the network. When it reaches to the output phase of the network where it achieves the target value, if it fulfils the set value , then the network maximizes the SINR mentioned in Eq.8 otherwise it is back to the LSTM network and again and updates its weights until it attains the final desired value where the graph of the corresponding signals coincides it with the original signal . At the end the required RMSE value is calculated form it. To do so the desired waveform of the raw signal is predicted by applying a LSTM algorithm of deep learning network.

## 5 | VEHICULAR TO INFRASTRUCTURE SCENARIO

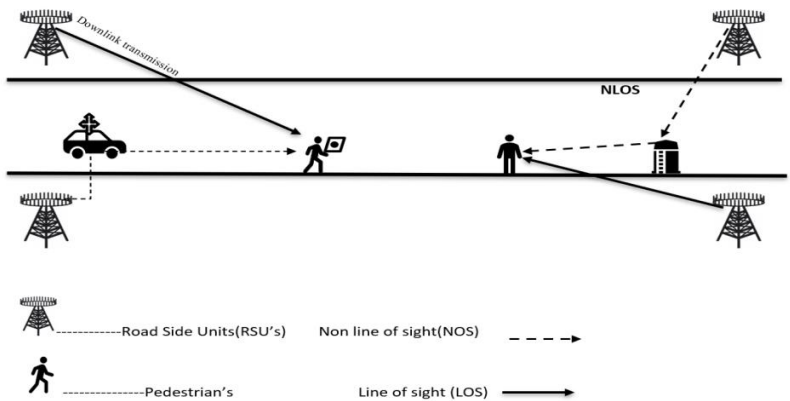

**FIGURE 2** The Proposed V2I Scenario

This paper employs adaptive beamforming forV2I communication by mitigation of interference signal with the help of MVDR beamformer technique.
In the V2I scenario mitigation of the interference is obtained by placing nulls in the signals. In this paper we have created a scenario and the assumed parameters according to [23] considering road side units (RSU's) with approximate height of 25m each corresponding to 700MHz bandwidth contains 64 antenna elements and the noise assumed to be 5dB.The inter-site distance between RSU's is 200m and the user element comprises of 64 antenna elements considering the same frequency measures with 23dBm power, the user may be pedestrian carrying mobile handset moving with speed of about 4km/hr, and vehicles with a speed of 30km/hr., thus the noise figure of user element is 7dB and due to the effect of multipath propagation, such that the signal being distorted and causes severe interference in case of non-line of sight (NLOS) as compared to line of sight (LOS).To mitigate the interference, an MVDR technique is applied by placing nulls at a certain angle by applying adaptive beamforming. The final beamformer leverages the regression capability of LSTM/Neural Network based regressor in which is required in 5G scenario to fulfil the demand of high SINR and throughput of system. This can be done by predicting the signal in advance by applying MVDR weights which automatically adjusts the beam in a desired direction by adjusting the angles(azimuthal) which plays a vital role in beamforming technique.

It is assumed that a rectangular pulse is incident on the uniform Linear array (ULA) at an azimuthal angle $45^0$ and $0°$ angle of elevation. Here we use the collect-planewave function of the ULA structure for the simulation of received pulse waveform from the incident angle. The signal has a matrix of 64 columns, where each column signifies the received signal at one of the array elements.

The next two section present a brief review of two popular adaptive schemes used in this study viz. LSTM and NAR based time series forecasting.

## 6 | OVERVIEW OF LONG SHORT-TERM MEMORY (LSTM)

To handle data dependent problems where there is a necessity to recognize the previous input for the further processing, RNN based LSTM networks is well suited to solve and predict the next state so that the structure of hidden layer is changed according to the output. [17]

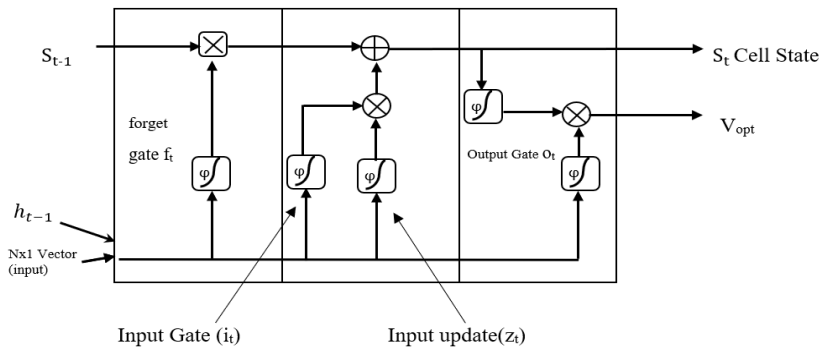

FIGURE 3  Basic Architecture of LSTM

As shown in Figure 3 the proposed LSTM structure provides a model that covers combination of different LSTM units, and consequently, contains different structure layer which is referred as input and output layer. As specifically, the LSTM neural networks comprise of multiple replicas of standard blocks of memory, and every component of block comprised a memory cell and three types of gates (input gate, output gate, and forget gate), as illustrated in Figure 3. The cell is the essential portion of LSTM memory and is viable for transfer of data into different steps of time. Each of the gates which corresponds to sigmoid function have option to transfer data which are accountable for state of cell by providing control and protection of it. The gate acts an input decides, which part of input is get processed further to update the cell. Correspondingly, the forget gate decides which portion of the previous cell state will be dumped, while the gate which governs, the output portion decides the next state will be output. At time step t for a memory block the $i_t, f_t$, and $o_t$ represents the input, forget, and output gates, respectively. Assume that $x_t$ and $h_t$ denotes the inputs and outputs at present time-instance, $h_{t-1}$ represents output at the previous time-instance, ψ represents the sigmoid activation function ,and ⊗denotes the hadamard product, the basic representations of the LSTM model are shown below

$$f_t = \sigma(w_{xf}x_t + w_{hf}h_{t-1} + b_f) \qquad (14)$$

$$i_t = \sigma(w_{xi}x_t + w_{hi}h_{t-1} + b_i) \qquad (15)$$

$$o_t = \sigma(w_{xo}x_t + w_{ho}h_{t-1} + b_o) \qquad (16)$$

$$\tilde{c}_t = \tanh\sigma(w_{xc}x_t + w_{hc}h_{t-1} + b_c) \qquad (17)$$

$$c_t = f_t \otimes c_{t-1} + i_t \otimes \tilde{c}_t \qquad (18)$$

$$h_t = o_t \otimes \tanh(c_t) \qquad (19)$$

Where b and w are the bias and weight metrices of the corresponding gates, after calculation of values of the respective gates, the steps to process further updating data through the structure of gate is divided into three steps.

In the starting step multiply value of the forget gate $f_t$ with previous state of cell $c_{t-1}$ chooses which portion of old cell state $c_{t-1}$ is discarded. Then, the cell state information is updated by multiplying of input gate value $i_t$ by the new content memory cell value $\tilde{c}_t$, At the final stage bymultiplying the output $o_t$ by the updated state of cell $c_t$ through a tanh function corresponds to the output value $h_t$. Now the final output $h_t$ and the cell state value $c_t$ will be passed to the next memory block atthe

$t + 1$ time-step [23].

## 7 | NON-LINEAR AUTO REGRESSIVE (NAR) MODEL

-: This well-known model uses the previous values of the time series to predict future output of the system. In most of the cases, applications of time series models are specified for higher deviations and momentary transient periods. Thus, this is difficult to implement time series model by using a liner model, therefore a nonlinear method should be suggested. A nonlinear autoregressive neural network,[24,25] is applied to time series forecasting, describes a discrete, non-linear, autoregressive model that given by

$$y(t) = h(y(t-1), y(t-2), \ldots y(t-p)) + \epsilon(t) \qquad (21)$$

discussed the concept of NAR method is applicable to predict the value of a data series y at time t, y(t), using the p past values of the series. The value of function h(.) is unknown in advance, and the optimization of this function through approximation is done by using the training of neural network by using the weights and neuron bias. Finally, the term $\epsilon(t)$ denotes the error of the approximation of the series y at time t.

The generalized architecture of a NAR network is shown in Figure 4. The p features y(t-1), y(t-2), . . ., y(t-p), represents the feedback delays. we can easily adjust the hidden layers and neurons per layer in the NAR. They are completely flexible and are optimizable through a trial-and-error technique to obtain the desired network performance. Levenberg-Marquardt backpropagation procedure (LMBP) is the most common learning method for the NAR network.[26-29] The second-order derivative was approximated by using the LMBP method with no need to compute the Hessian matrix, thus increases the training speed.

The Hessian matrix can be approximated if the performance function has the form of a sum of Squares (generally used in the feed forward neural network training), then the hessian matrix can be represented as

$$H = J^T J \tag{22}$$
$$G = J^T e \tag{23}$$

Equations (22) and (23), $J$ denotes the Jacobian matrix constitutes the first derivatives of the network errors with respect to the biases and weights, and denotes the error of the network in all training samples.

The algorithm (LMBP) uses this method in the Newton-like update described in Equation 24.

$$\chi_{k+1} = \chi_k [J^T J + \mu I]^{-1} J^T e \tag{24}$$

The performance function used to calculate the jacobian matrix used in this technique is mean of the sum of the square error and it is discussed in Eq.25.

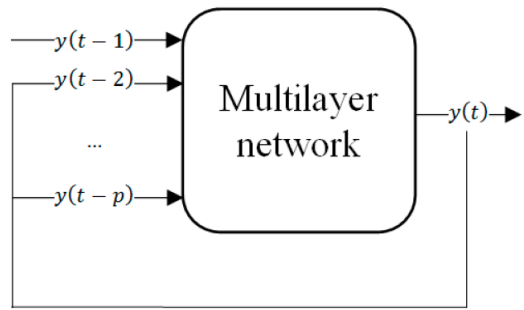

**FIGURE 4**   General Architecture of NAR

## 8 | DATASET DESCRIPTION

The dataset contains an Nx1 vector as an input which represents the beamforming weights of N rows consists of 361 values and one column was considered as a beamformer output. The transmitted signal having complex value is arriving on an array of 64 elements from the direction of azimuthal angle $45^0$ and $0^0$ elevation. To compute the mean square error weights of the received signal, noise is added to it, thus the data is generated based on the scenario discussed in section 5. Here the rectangular coordinate system is used while generating the weights and then desired signal is shown in fig 3.1(a). Each data points contain complex values, represented by either a two-dimensional coordinate according to the requirement of spatial geometry for V2I applications. The use of the training data of interference for a LSTM network partitioned into three parts: (i)An input layer which is fully connected sequences of positions of the moving vehicle and pedestrian, such that every value in the sequence is correlated to a multidimensional vector(ii) The processed sequence is sent to the foremost part of the interference model. (iii) A fully connected output layer maps the output of the last LSTM layer at each time-step. The purpose of training is to reduce the loss function, which can be either normalize mean square error (NMSE) when the input Nx1 dimensional vector which contains the complex weights $w_i$. The training signal used to train the network is considered as the received signal impinging on the array. We can train a sequence-to-sequence regression LSTM network to predict the values of future time steps, where the responses are the training sequences with values shifted by one time step. That is, at each time step of the input sequence, the LSTM network learns to predict the value of the desired signal in next time step.

Here an LSTM network forecasted the number of beamformed signals as an output with given the number of beamforming weights. The output is a cell array, where each element is a single time step. Reshape the data to be a row vector and then split the data in to training and testing phase in 80-20 % ratio after that fit the dataset into the network properly and avoid over fitting, to standardize the training data it is set to have zero mean and unit variance. At prediction time we must standardize the test data using the same parameters as the training data. To train the network we used the specified training options as train Network in MATLAB 2020a.

We are showing that the outcome of the training process is used to train the beamformed signal for the prediction for three cells: since they are deployed in to three different regions, the RSU's presents distributed interference profiles in terms of number of iterations and amplitude of signal. We can see that the prediction for number of iterations with the oscillating nature of the interference, in this scenario, the forecast is one-step ahead, implying that we are using a set of past values to estimate the direction of arrival for the next time slot. Here, the predictions are better than earlier calculations when it is updating the state of network with the observed parameters than the predicted values.

So, to evaluate the performance of the proposed architecture, we calculate the interference on average basis thus it is a synthetic generated data which is not calculated at particular instances thus we choose the mean square error (MSE) of this quantity. For the ease of calculation, it is then further normalized. Which is the metric to measure the accuracy of the prediction algorithm, and is given as

$$\text{RMSE} = \sqrt{\frac{1}{\bar{x}} \frac{\sum_{t=1}^{M} (\tilde{x}_t - x_t)^2}{M}} \tag{25}$$

Here M is number of instances of total values, $\tilde{x}_t$ and $x_t$ is the predicted value at time t and $\bar{x}$ denotes its mean.

## 9 | RESULTS AND DISCUSSION

In this section we discuss the performance of the two networks in terms of root mean square error and normalized power in the forecast phase .

Fig. shows the comparison of root mean square error values for LSTM model with the NAR model. The effect of prominent array parameters on error performance of the predictor have been depicted in Fig. . This section also provides valuable insights in terms of optimal array parameters resulting in a near optimal beamformed signal and improved performance of the network. by predicting the interference signal through proper training of weights and shows the error given by the network.

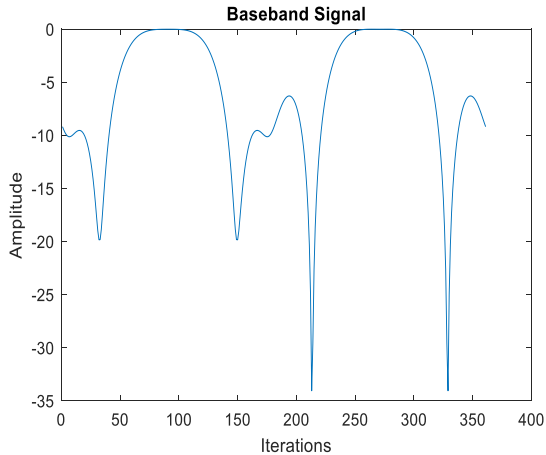
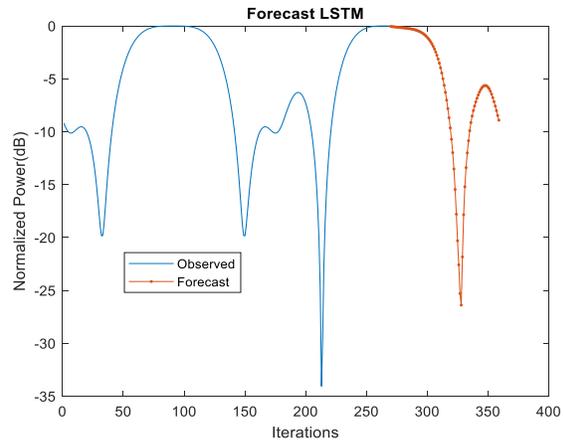

FIGURE 5.1 – Baseband Signal    FIGURE 5.2 Beamformed Signal

A comparative inspection of Figs. 5.2 and 5.3 reveals that the beamformed signal forecasted using NAR has higher similarity to the baseband signal than that forecasted using LSTM when the incident angle parameter of the received signal was varied. This is evident from a sharper dip in the normalized power of the beamformed signal in Fig. 5.3. A lower degree of similarity in the forecasted beam pattern in the case of LSTM can be attributed to a more complex training and update procedure, inherent to LSTM module. It was also observed that the similarity between the forecasted signal and the baseband signal was directly correlated with the RMSE obtained in the testing phase. Henceforth, for the sake of brevity only RMSE values have been depicted.

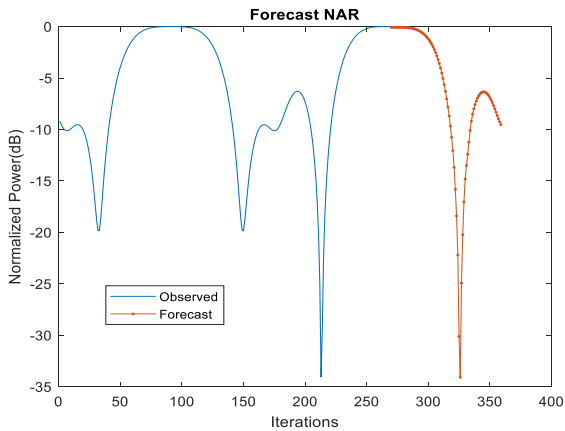
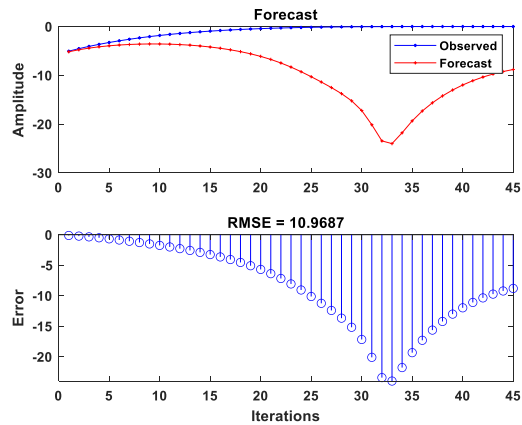

FIGURE 5.3 Forecasted NAR signal    FIGURE 5.4 Training Signal Before Forecast

Fig. 5.3 depicts a part of it which was forecasted by the NAR regressor with a small enough test phase error.

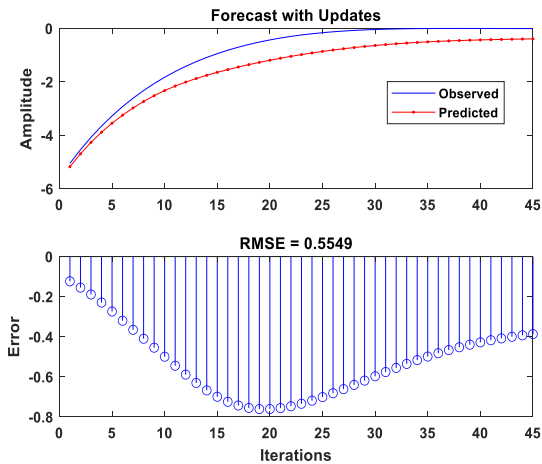

**FIGURE 5.5 - Forecasted Signal After Training**

We can observe from Figs. 5.4 and 5.5 the effect of weight update on the prediction performance of LSTM network. The forecasted signal power (normalized) before and after update are depicted in the top part whereas, the bottom part shows the effect of update on test phase RMSE which is more than several orders of magnitude less in the latter case.

The forecasted results shown in Fig. 5.1-5.5 corresponds to the case when received signal incident angle was varied from 40 to 100 degrees keeping all other parameters fixed.

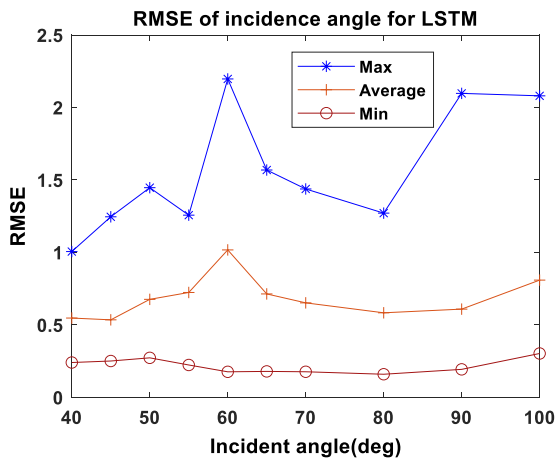
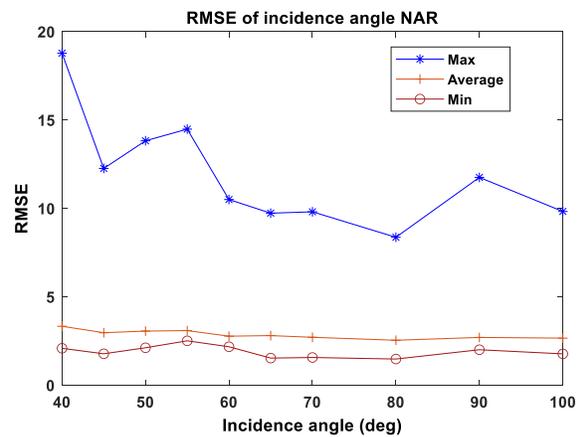

**FIGURE 6.1 – Comparison of RMSEs of regressor for incident angle parameter only for LSTM**  **FIGURE 6.2 – Comparison of RMSE for NAR only**

The RMSEs of LSTM and NAR regressor are depicted in Fig. 6.1 when incident angle was varied from 40 to 100 degrees as the primary input for the beamformer. It can be observed that the obtained RMSEs of both regressor are comparable. Since, the experiments were repeated several times with k-fold cross validation scheme (k=10), in Fig. 6.1, average value of the RMSE obtained in k-folds has been plotted.

For the reader's ready reference, Fig. 6.2 has also been plotted depicting maximum, minimum and average values of the RMSE obtained using LSTM network.

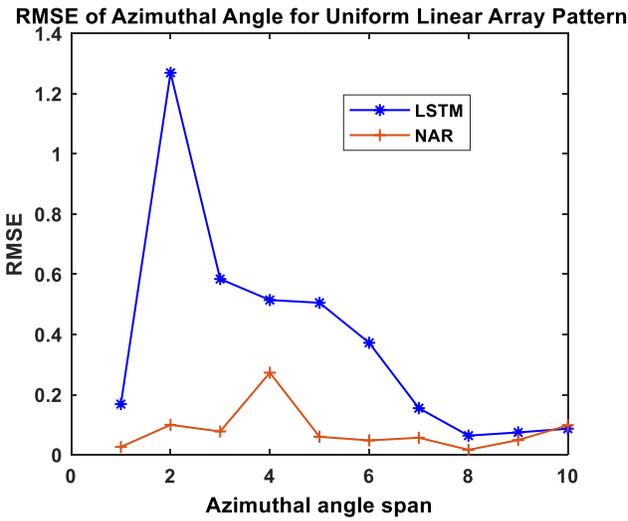

FIGURE 7.1 – Comparison of RMSE

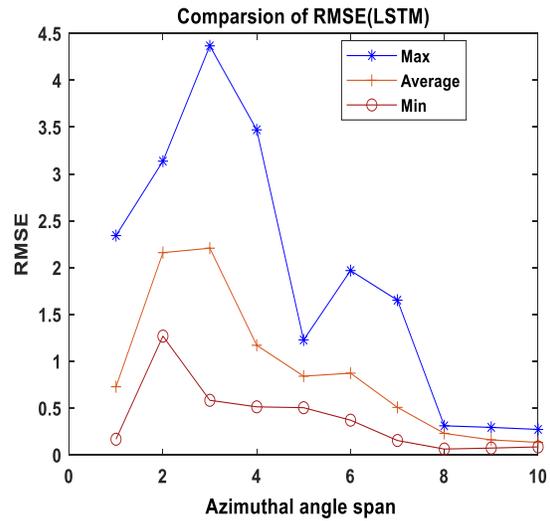

FIGURE 7.2-Comparison of RMSE only for azimuthal angle for LSTM

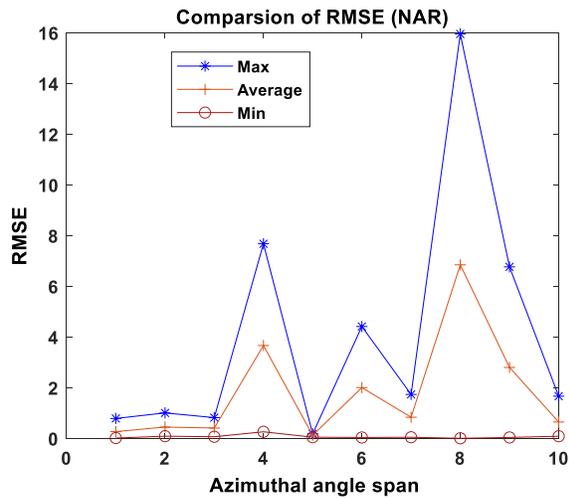

FIGURE 7.3 RMSE Comparison for NAR

Similarly, the RMSEs obtained when primary input for the beamformer was azimuthal angle have been shown in Figs. 7.1 to 7.3. In this case NAR exhibits better performance than LSTM. In the figures the angle spans were coded from 1 to 10 using the scheme shown in table 1.

| Angle Span | Designated As |
|---|---|
| -30 to +30 | 1 |
| -45 to +45 | 2 |
| -60 to +60 | 3 |
| -60 to +60 | 4 |
| -70 to +70 | 5 |
| -80 to +80 | 6 |
| -90 to +90 | 7 |
| -100 to +100 | 8 |
| -110 to 110 | 9 |
| -120 to +120 | 10 |
| 130 to +130 | 11 |

**TABLE 1- Designation for Azimuthal Angles**

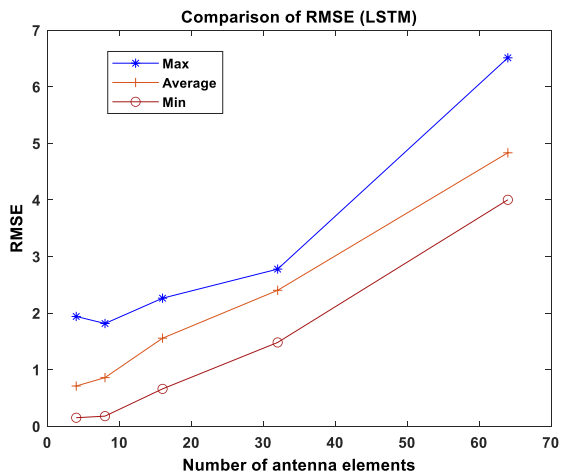
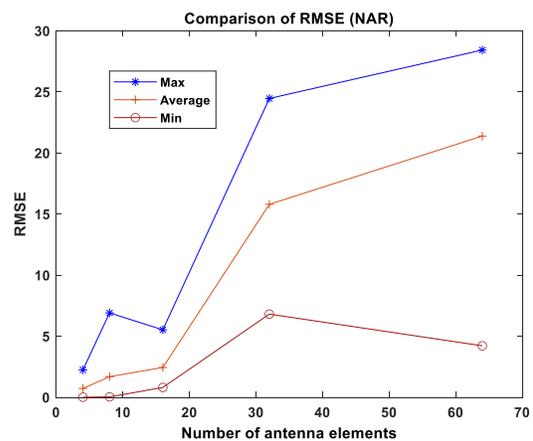

**FIGURE 8.1 RMSE comparison for LSTM**     **FIGURE 8.2 RMSE comparison for NAR**

Perhaps the most significant advantage of LSTM regressor can be observed from Figs. 8.1 and 8.2 where number of antenna elements are varied from four to sixty four. Here, RMSE of LSTM is at least an order of magnitude less than that of the NAR regressor. It is likely, that complexity incurred by increasing the number of antenna elements is taken care of by the LSTM module, whereas, the NAR module underperforms since there is no provision there for learning dependencies. Therefore, in a realistic scenario, where, the antenna/sensor elements are large in number, it would be advisable to employ an LSTM based forecast stage for beamforming.

# 10 | CONCLUSION

Based upon the results presented in the previous section, we can conclude that neural network based regressor can be employed to successfully forecast beamformed signals. Data generated by MVDR beamformers (or possibly any other conventional method) can be used for training. Further, LSTM holds a lot of promise for forecasting beamformed signals since there are several short and long term dependencies in a V2I scenario that needs to be learned by the regressor. For the general V2I scenario considered in this work LSTM outperforms NAR based regressor especially when the number of antenna elements is large. There is scope for future work in optimizing the architecture of LSTM module with an aim to minimize computational complexity while at the same time improving the beamforming forecasts.